This brief note in response to Wojciech Zurek's article "Quantum Darwinism, classical reality, and the randomness of quantum jumps" (Physics Today, October 2014, page 44) points out extant rebuttals in the literature to some of the author's key claims.

On the first page of his arXiv.org version of the article (http://arxiv.org/abs/1412.5206), Zurek states that

> decoherence selects preferred pointer states that survive interaction with the environment. They are localized and effectively classical. They persist while their superpositions decohere. Decoherence marks the border between quantum and classical, alleviating concern about flagrant . . . manifestations of quantumness in the macroscopic domain.
>
> Here we consider emergence of "the classical" starting at a more fundamental pre-decoherence level, tracing the origin of preferred pointer states.

However, the idea that preferred pointer states naturally "emerge" from the quantum level has been refuted in the published literature, in particular in a paper[1] showing that "classical" pointer states do not emerge unless a key aspect of classicality has been tacitly assumed from the beginning. In other words, the "quantum Darwinism" program is fatally circular.

The assumption generating the circularity usually takes the form of a predesignated system that is considered separable from its environment (the same assumption appears in the discussion about "information flow" in Zurek's article). The system acquires decohered observable (or "stable") states because it is presumed to be distinguishable from a designated set of environmental subsystems, all assumed as having random phases with respect to each other and the system. The only correlations between the system and the environmental subsystems are assumed to be established via the designated Hamiltonians.

However, in the absence of this sort of predesignated partitioning of all degrees of freedom into the system and its surroundings—measurement apparatus, environment, and so forth—where the initial phase of each subsystem is random, the desired decoherence and emergence of pointer states, or einselection, do not follow. The partitioning is inevitably based on what a human observer would be able to identify and measure, such as distinguishable atoms and molecules, so the account is dependent on assuming the classical realm of the observer.

The need for this pre-partitioning fatally conflicts with the claim that classicality emerges naturally from the quantum realm: Absent a pre-partitioning of all degrees of freedom into uncorrelated systems of interest, classical pointer states do not emerge. With unitary-only dynamics (lacking non-unitary collapse), the quantum realm does not have any *a priori* preference for the assumed uncorrelated degrees of freedom. On the contrary, a unitary-only evolution would typically begin with a maximally entangled universal state.

As I noted in reference 1, one can observe the decoherence process experimentally, but that observation does not demonstrate that einselection occurs in a unitary-only dynamics or that classicality emerges in such a dynamics. A key missing ingredient in quantum Darwinism is some real physical randomization process, such as collapse, that could create an observer-independent, physical partitioning at the quantum level. Without such a process, quantum Darwinism contains the same kind of logical circularity as Boltzmann's *H*-theorem, which attempted to derive irreversible thermodynamic laws from reversible laws. Boltzmann inadvertently smuggled in irreversibility by assuming molecular chaos; quantum Darwinists smuggle in classicality via their partitioning of the universe into distinguishable systems of interest that interact with mutually randomized environmental subsystems.

In his article, Zurek does not address this refutation of "quantum Darwinism" and repeats the refuted claims and circular assumptions. Similar criticisms of quantum Darwinism's presumption of classically distinguishable primordial systems, raised by Jasmina Jeknić-Dugić, Miroljub Dugić, and collaborators,[2–4] and Chris Fields's concern about the division of primordial degrees of freedom into a classically distinguishable system and environment[5] have also not been countered by proponents of quantum Darwinism.

*I acknowledge and thank Miroljub Dugić for helpful comments.*

**Ruth E. Kastner**

*(rkastner@umd.edu)*

*University of Maryland*

*College Park*